# Impact of non-Hemiticity on modal strength and correlation in transmission through random open cavities

Matthieu Davy[1] and Azriel Z. Genack[2]

[1]Univ Rennes, CNRS, Institut d'Électronique et de Télécommunications de Rennes, UMR–6164, F-35000 Rennes, France
[2]Department of Physics, Queens College and Graduate Center of the City University of New York, Flushing, New York 11367, USA

The nonorthogonality of eigenfunctions over the volume of non-Hermitian systems determines the nature of waves in complex systems. Here, we show in microwave measurements of the transmission matrix that the non-Hermiticity of open random systems leads to enhanced modal excitation and strong correlation between modes. Modal transmission coefficients reach values comparable to the dimensionless conductance which may be much larger than unity. This is accompanied by strong negative correlation between modal speckle patterns ensuring that net transmission is never larger than the incident power.

Excitation of and transport through a complex medium reflect the character of the eigenstates of the wave equation. In quantum systems these are referred to as energy levels, while for classical waves, these are called quasi-normal modes, or simply modes. Though it is generally not possible to solve for the eigenvalues of the Hamiltonian in large complex systems, resonances in the complex frequency-plane correspond to the poles of the scattering matrix $S$ and can be extracted from measurement of the spectrum of its elements [1-7]. The statistics of level spacing and width [8-12] and the scaling of conductance or transmission [13-15] have been studied for many years but less is known about the degree of correlation between eigenfunctions of overlapping modes in non-Hermitian systems and the degree to which they are excited by incident radiation.

A single mode may be excited when the sample is illuminated on resonance with a spectrally isolated mode. However, as the coupling to the exterior and internal dissipation increase, modes broaden and overlap spectrally. In such non-Hermitian systems, the eigenfunctions are nonorthogonal [1, 5, 16-19]. The non-orthogonality of eigenfunctions leads to the existence of exceptional points in systems that incorporate both gain and loss but in which parity-time symmetry is preserved [20, 21] and to the enhancement of the linewidth and spontaneous emission rates in laser resonators [22-24]. Its impact has also been explored in localized plasmonic surfaces [25], dielectric microcavities [26, 27] and chaotic systems with small perturbations [28, 29]. However, direct demonstrations of the nonorthogonality of modes in open systems and its influence on the statistics of modal excitation have remained a challenge.

Microwave measurements in the region of moderate modal overlap have shown that modal strength in transmission may be enhanced in isolated cases and interference between modal speckle patterns tends to suppress transmission below the incoherent sum of modal contributions [3, 30, 31]. Determining the statistics of individual modes and the degree of interference between modal speckle patterns is central to describing the propagation of waves and to controlling the flow of radiation by shaping the incident wavefront in photonic [32-34] and plasmonic systems [4, 6, 35].

Here we demonstrate the relation between the correlation of eigenfunctions over the volume of the sample and the correlation of modal components of the transmission matrix (TM) $t(\omega)$ in non-Hermitian systems. This leads to a systematic linkage of enhanced modal transmission and destructive interference among correlated eigenfunctions of overlapping resonances. The degree of modal overlap in random media is varied by changing the degree of disorder or the openness of the sample boundaries. The modal overlap may be characterized by the Thouless number $\delta$, which is the ratio of the average linewidth to the average level spacing $\delta = \langle \Gamma_n \rangle / \Delta \omega$. $\delta$ reflects the degree of spatial localization since tightly localized modes have narrow linewidths since they couple weakly to the surroundings through the sample boundaries [10, 36]. The Thouless number is equal to the average of the dimensionless conductance $\delta = g$, which similarly falls as modes are more strongly localized and transport is suppressed. For classical waves, the dimensionless conductance corresponds to the average transmittance, $g = \langle T(\omega) \rangle$, where, $T(\omega) = \Sigma_{ab} |t_{ba}(\omega)|^2$ is the sum over flux transmission coefficients between all incoming and outgoing channels, $a$ and $b$, respectively [10, 14, 36, 37].

We measure the TM of a multichannel two-dimensional random system (see Fig. 1(a)). The disordered aluminum cavity of height $H = 8$ mm, width $W = 250$ mm and length $L = 500$ mm supports a single transverse mode in the vertical direction. The randomly positioned scattering elements are 6-mm-diameter aluminum cylinders. The TM which is the part of the scattering matrix associated with the transmission coefficients between incoming and outgoing channels on the left and right side of the sample, respectively, is measured in the microwave range between two linear arrays of $N = 8$ antennas that are coaxial to waveguide adapters [31]. Spectra of each transmission coefficients of the TM are successively obtained using two electro-mechanical switches and the two ports of a vector network analyzer (VNA). The openings of the system are fully controlled by the antennas [38] but strong internal reflection may appear at the interfaces due to the metallic region surrounding each coupler in comparison to a waveguide which is fully open on both ends.

The eigenfunctions are the complex right $|\phi_n\rangle$ and left $\langle \varphi_n |$ eigenvectors of the wave equation with outgoing boundary conditions, $(\Delta + k^2)\psi(r) = 0$. They form two complete bi-orthogonal sets satisfying the orthogonality relations $\langle \varphi_n | \phi_m \rangle = \delta_{nm}$ and are associated with complex eigenvalues $\widetilde{\omega}_n = \omega_n - i\Gamma_n/2$, where $\omega_n$ is the central frequency and $\Gamma_n$ is the linewidth. For systems with time-reversal symmetry, the eigenvectors are related by the transpose $\langle \varphi_n | = (|\phi_n\rangle)^T$ and the matrix of eigenfunctions $\phi_n$ is normalized by $\phi^T \phi = I$. The expansion of the scattering matrix in terms of quasi-normal modes is then $S = I - iW^T[\omega - \widetilde{\Omega}]^{-1}W$. The matrix $W$ of vectors $W_n$ is the projection of eigenfunctions $\phi_n$ onto the channels of the sample and $\widetilde{\Omega}$ is the diagonal matrix of eigenvalues $\widetilde{\omega}_n$.

The resonances $\widetilde{\omega}_n$ and the projection of eigenfunctions on the interfaces are found in an analysis of spectra of the TM as a superposition of modal TMs (MTMs) [31]

$$t(\omega) = -i\Sigma_n \frac{W_{Rn} W_{Ln}^{\mathrm{T}}}{\omega - \widetilde{\omega}_n}. \tag{1}$$

$W_{Ln}$ and $W_{Rn}$ are the components of the $W_n$ vector associated with the left and right sides of the sample, respectively. The modal analysis is performed using the Harmonic Inversion technique to extract the central frequencies and linewidths from spectra of transmission coefficients [2, 31]. Each modal transmission coefficient giving the vectors $W_{Ln}$ and $W_{Rn}$ is then reconstructed from a fit of the corresponding transmission coefficient spectrum as a superposition of Lorentzian lines. Equation (1) shows that the MTM $t_n = -iW_{Rn}W_{Ln}^{\mathrm{T}}$ is of unit rank and $W_{Ln}$ and $W_{Rn}$ correspond to the modal speckle patterns of the nth mode. The validity of the expansion of the TM into MTMs is confirmed within the accuracy of the modal decomposition by the finding that the ratio between the two first eigenvalues of each MTM is higher than $10^2$. The vectors $W_{Rn}$ and $W_{Ln}$ are then extracted from the singular value decomposition of each MTM.

Figure 1(b) shows the spectrum of the transmission through the antennas determined from $T_c(\omega) = 1 - |\langle S_{cc}(\omega) \rangle|^2$, where $\langle S_{cc}(\omega) \rangle$ is the mean reflection parameter at each antenna. We carry out measurements in three ensembles with moderate modal overlap in frequency ranges in which: (1) the antennas are weakly coupled to a sample ($T_c \sim 0.09$) with 30 cylinders contained within a cavity, for which $\delta \sim 0.5$; and (2,3) the antennas are strongly coupled to a sample ($T_c \sim 0.98$) and the disorder is strong. For the samples with 280 and 200 scatterers, $\delta \sim 1.2$ and $\delta \sim 2.01$, respectively.

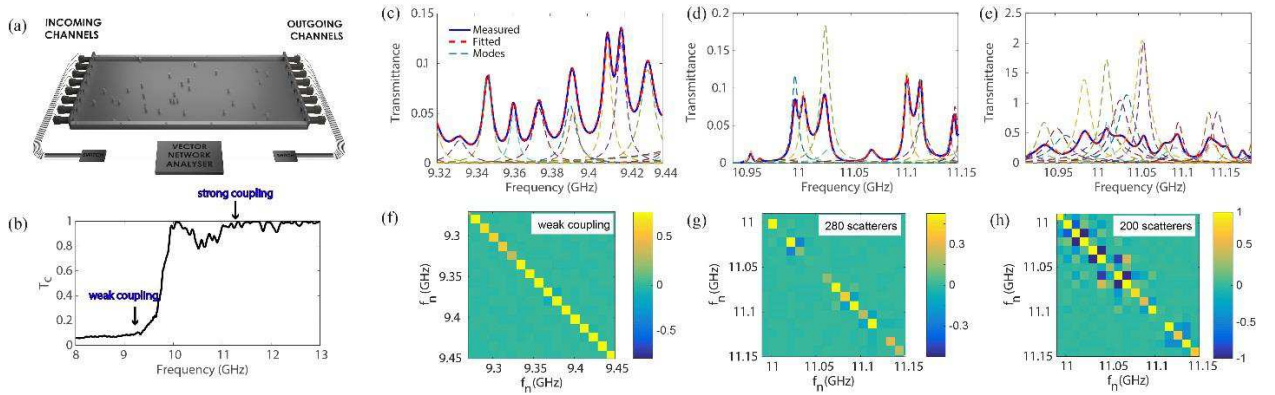

FIG. 1 (double-column figure): (a) Experimental setup. (b) Transmission through the antennas as a function of frequency. The arrows indicate the frequency ranges corresponding to weak and strong coupling of the antennas. (c-d) Measured transmittance $T(\omega) = \Sigma_{a,b}|t_{ba}(\omega)|^2$ and the underlying modal structure for (c) weak coupling to a sample with 30 cylinders and (d) for strong coupling to a sample with 250 cylinders and (e) 200 cylinders. (f-h) Corresponding real part of the overlap matrix. The modal strength and negative correlation between neighboring modes are seen in the increase of the diagonal and off-diagonal terms, respectively, with increasing modal overlap from (f) to (h).

Spectra of the transmittance $T(\omega) = \Sigma_{ab}|t_{ba}(\omega)|^2$ and the modal strengths in transmission $T_n(\omega) = \|W_{Rn}\|^2\|W_{Ln}\|^2/|\omega - \widetilde{\omega}_n|^2$, are shown in Fig. 1(c-e). In the vicinity of the central frequency of an isolated mode, peaks in transmission spectra correspond to resonances with individual modes, $T(\omega) \sim T_n(\omega)$. But as modal overlap increases with increasing coupling of the antennas to the sample, the strength of modes on resonance $T_n = T_n(\omega_n)$ is enhanced and may exceed unity for $\delta = 2.01$.

We now show that measurement of the TM makes is possible to explore statistics of the overlap matrix $O_{mn} \equiv \langle\varphi_m|\varphi_n\rangle\langle\phi_m|\phi_n\rangle = \left[\int dr \phi_m^*(r)\phi_n(r)\right]^2$ which characterizes the correlation between eigenvectors over the volume [17, 19, 39]. The diagonal elements of $O_{mn}$ are equal to the Petermann factor, $K_n \equiv O_{nn}$, which is a measure of the degree of complexness of the eigenfunctions [40-43]. The Petermann factor characterizes the excess spontaneous emission for laser cavities and governs the linewidth of lasing modes [22-24, 44]. For small modal overlap, the eigenfunctions coincide closely with the real eigenfunctions of the closed system so that $K_n \sim 1$. However, $K_n$ increases as the coupling of the sample to its surroundings increases [41] and can have values exceeding one thousand [23].

The off-diagonal elements of $O$ give the degree of correlation between eigenfunctions. Since the eigenfunctions are complete, $\Sigma_m O_{nm} = 1$ [17], the enhancement of $K_n$ with increasing $\delta$ implies that non-vanishing and negative-on-average correlation $O_{n\neq m} < 0$ between overlapping eigenfunctions, in contrast to the orthogonality in Hermitian systems.

A direct probe of the overlap matrix would require a non-invasive scan of the spatial profile of the eigenfunctions inside the sample, which is almost impossible in most cases. However, the correlation of eigenfunctions is expressed in the correlation of their projections onto the coupling channels [5, 45, 46]

$$O_{mn} = -\frac{\left(W_m^\dagger W_n\right)^2}{(\widetilde{\omega}_n - \widetilde{\omega}_m^*)^2} \tag{2}$$

For $m = n$, Eq. (2) gives the relation between the linewidths and the coupling vectors, $K_n = \|W_n\|^4/\Gamma_n^2$ [45]. In principle, Eq. (2) makes it possible to obtain the degree of nonorthogonality of eigenfunctions from the decomposition of the scattering matrix into a sum of Lorentzian lines. Because we measure the TM rather than the scattering matrix, it is not possible to find the vectors $W_{Ln}$ and $W_{Rn}$ separately, only the MTMs $t_n = -iW_{Rn}W_{Ln}^{\mathrm{T}}$ can be calculated. The relative phase and magnitudes of the vectors $W_{Ln}$ and $W_{Rn}$ are unknown. We compute the "transmission overlap matrix"

$$\tilde{O}(\tilde{\omega}_m, \tilde{\omega}_n) = -\frac{4(W_{Rm}^\dagger W_{Rn})(W_{Lm}^\dagger W_{Ln})}{(\tilde{\omega}_n - \tilde{\omega}_m^*)^2} = -\frac{4\mathrm{Tr}(t_m^\dagger t_n)}{(\tilde{\omega}_n - \tilde{\omega}_m^*)^2} \qquad (3)$$

The diagonal elements are $\tilde{O}_{nn} = \tilde{O}(\tilde{\omega}_n, \tilde{\omega}_n) = T_n$. $\tilde{O}(\tilde{\omega}_m, \tilde{\omega}_n)$ and $O(\tilde{\omega}_m, \tilde{\omega}_n)$ are equal for extended states in the limit $N \gg 1$ for which $\|W_{Rn}\| \sim \|W_{Ln}\|$ [46]. We also compensate the impact of absorption on the operator $\tilde{O}$. Since the linewidths $\Gamma_n$ are broadened by absorption, we replace in the denominator of Eq. (3) $\tilde{\omega}_n - \tilde{\omega}_m^*$ by $\tilde{\omega}_n - \tilde{\omega}_m^* + i\Gamma_a$, where $\Gamma_a$ is the homogeneous absorption rate, with $\Gamma_a \sim 3.6$ MHz in the low frequency range with weakly coupled antennas and $\Gamma_a \sim 4$ MHz in the higher frequency range with strongly coupled antennas [46].

The real parts of $\tilde{O}(\tilde{\omega}_1, \tilde{\omega}_2)$ are shown in Fig. 2(f-h) for the three samples studied. When the antennas are weakly coupled, the overlap matrix $\tilde{O}$ is seen to be close to diagonal as it would be for a closed system. Uniform losses that broaden the linewidths indeed do not alter the orthogonality of eigenfunctions. For the case of strong coupling and $\delta \sim 1.2$, the transmission overlap matrix is also mostly diagonal. However, when two resonances overlap enhanced diagonal and negative off-diagonal elements are observed, as seen, for instance, for the two resonances with central frequencies around 11.03 GHz. For the sample with the smaller number of scatterers and hence shorter mode lifetime and larger mode linewidth, $\delta \sim 2.01$, the diagonal part for most modes increases while the off-diagonal terms become more negative.

To carry out measurements on a random ensemble, we move a 10 mm-diameter magnet along a line within the medium in steps of $\lambda/2 = 12.5$ mm, where $\lambda$ is the wavelength at 12 GHz. The magnet within the sample is moved by the force of a second magnet above the top plate of the cavity. We find resonances and associated modal coefficients for more than a thousand modes in 40 realizations of two ensembles: 1) weakly coupled antennas with 30 scatterers giving $\delta = 0.5$, and 2) strongly coupled antennas with 280 scatterers. Since no scatterers are positioned along the line of motion of the magnet, $\delta$ for this ensemble is increased to $\delta = 1.52$ from $\delta = 1.2$, for the sample where there is no excluded volume for scatterers.

We first explore the degree of correlation between different modes, the off-diagonal elements of $\langle \tilde{O}(\tilde{\omega}_1, \tilde{\omega}_2) \rangle$. The average $\langle \tilde{O}(\tilde{\omega}_1, \tilde{\omega}_2) \rangle$ is shown in Fig. 2 as a function of the complex shift between two resonances $|\tilde{\omega}_1 - \tilde{\omega}_2|$ normalized by $\Delta\omega$. $\langle \tilde{O}(\tilde{\omega}_1, \tilde{\omega}_2) \rangle$ is seen to be negative with a magnitude which decreases with $|\tilde{\omega}_1 - \tilde{\omega}_2|$. The magnitude of all elements are seen to be stronger for strongly coupled antennas as a consequence of greater nonorthogonality.

Chalker and Mehlig predicted that the eigenvector correlator of $M \times M$ non-Hermitian random matrices of the Ginibre complex Gaussian ensemble is given by [17, 19, 39]

$$O(\tilde{\omega}_1, \tilde{\omega}_2) \sim -\frac{1}{|\delta\tilde{\omega}|^4}[1 - (1 + |\delta\tilde{\omega}|^2)\exp(-|\delta\tilde{\omega}|^2)], \qquad (4)$$

where $\delta\widetilde{\omega} = \sqrt{M}(\widetilde{\omega}_1 - \widetilde{\omega}_2)$ is essentially the complex spacing between resonances. This result was confirmed for non-Hermitian random matrices describing the statistical properties of resonances in open chaotic cavities [18]. To compare theoretical and experimental results, the complex spacing is normalized by the level spacing and a scale factor $a$ of order of unity, $|\delta\widetilde{\omega}| = |\widetilde{\omega}_1 - \widetilde{\omega}_2|/(a\Delta\omega)$. A good fit of experimental results for $\widetilde{O}(\widetilde{\omega}_1, \widetilde{\omega}_2)$ in Fig. 2 is obtained using Eq. (4). The power law tail of $\widetilde{O}(\widetilde{\omega}_1, \widetilde{\omega}_2)$ as $|\widetilde{\omega}_1 - \widetilde{\omega}_2|^{-4}$ is confirmed in the inset when $|\widetilde{\omega}_1 - \widetilde{\omega}_2| > 2\Delta\omega$.

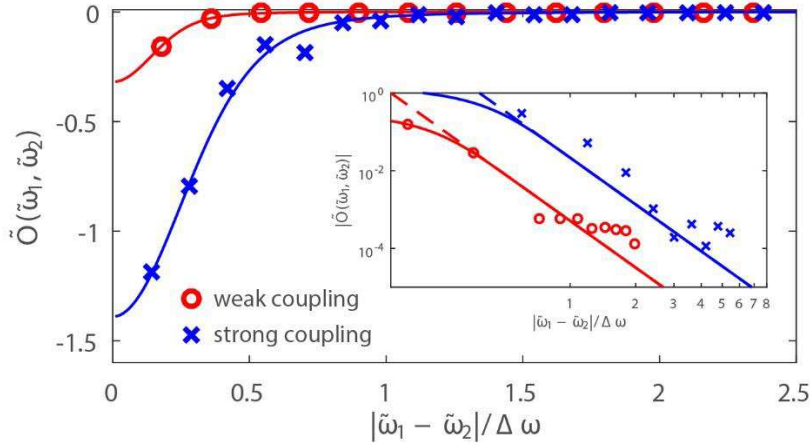

FIG. 2: Measured correlator $O(\widetilde{\omega}_1, \widetilde{\omega}_2)$ plotted as a function of $|\delta\widetilde{\omega}| = |\widetilde{\omega}_1 - \widetilde{\omega}_2|/(a\Delta\omega)$, where the scale factor is $a = 0.15$ and $a = 0.52$ for the two ensembles with weakly (red circles) and strongly (blue crosses) coupled antennas. The lines are fits to theoretical prediction given in Eq. (4). (inset) $|O(\widetilde{\omega}_1, \widetilde{\omega}_2)|$ in logarithmic scale is seen to scale as $|\delta\widetilde{\omega}|^{-4}$ (dashed lines) for $|\delta\widetilde{\omega}| > \Delta\omega$.

The distribution of the modal strengths in transmission which are the diagonal elements $T_n = \widetilde{O}_{nn}$ is shown in Fig. 3(a). The distribution $P(T_n)$ extends between 0 and unity and is peaked near $T_n = 0$. Values as large as $T_n = 4$ are found in the tail of the distribution for $T_n > 1$.

We now analyze the variation of the enhancement of $T_n$ with $\delta$. In the absence of absorption, $\sqrt{K_n}\Gamma_n = \|W_n\|^2$ so that $T_n$ can be expressed as the product of two terms of different origin

$$T_n = \frac{4\|W_{Rn}\|^2\|W_{Ln}\|^2}{(\|W_{Rn}\|^2+\|W_{Ln}\|^2)^2} K_n \equiv C_n K_n. \qquad (5)$$

Here, $C_n = \frac{4\|W_{Rn}\|^2\|W_{Ln}\|^2}{(\|W_{Rn}\|^2+\|W_{Ln}\|^2)^2}$, with $C_n \leq 1$, is the coupling asymmetry for the nth eigenfunctions between the left and right boundaries, which reflects the spatial pattern of the eigenfunctions within the sample.

We investigate the statistics of $T_n$ in random media in the crossover from diffusion to localization using the tight-binding Hamiltonian (TBH) model [47, 48]. The Hamiltonian of the closed random system of dimension $(NL)\text{x}(NL)$ is $H_0 = \Sigma_n |n\rangle A_n \langle n| + \Sigma_n (|n\rangle\langle n+1| + |n+1\rangle\langle n|)$. The on-site potential $A_n$ is independently and uniformly distributed on the interval $[-A/2, A/2]$. Each lead is modeled by a 1D semi-infinite TBH so that the effective Hamiltonian is given by $H_{\text{eff}} = H_0 - e^{ik} V V^T$, where $V$ is an $M$x$N$ matrix with elements equal to unity for sites to which the leads are attached and zero elsewhere [47, 48]. The wavevector $k$ is $\pi/2$ in the center of the band at $\omega \sim 0$. TBH simulations with $L = 200$ and $N = 20$ are carried out for different values of $A$ with $g$ ranging from 5.6 to 0.02.

For localized waves, $g \ll 1$, the distribution $P(T_n)$ shown in Fig. 3(c) is bimodal with peaks at $T_n = 0$ and $T_n = 1$. This is a consequence of the bimodal distribution of asymmetry factors $C_n$ of spatially localized modes. The bulk of the distribution can be explained by considering the coupling to a localized eigenstate exponentially peaked at $x_0$ in the sample with localization length $\xi$. The strength of the eigenfunctions at the left and right interfaces is given by $\|W_{Ln}\|^2 \sim e^{-\frac{x_0}{\xi}}$ and $\|W_{Rn}\|^2 \sim e^{-\frac{L-x_0}{\xi}}$. Hence, Eq. (5) gives, $C_n \sim \cosh^{-2}(\frac{L-2x_0}{2\xi})$. Assuming a uniform distribution of $x_0$ between 0 and $L$ leads to a bimodal distribution of $C_n$ [49]. This is in agreement with the formula proposed for isolated peaks in the transmission spectrum of 1D samples using a resonator model associated with effective cavities of length $\xi = \ell$ [49, 50]. When transmission is dominated by a single mode, $K_n \sim 1$ and $T_n \sim C_n$, but modes may occasionally overlap even in an ensemble in which $g < 1$ [51]. $K_n$ may then be large and $T_n$ can significantly exceeds unity to produce a tail in $P(T_n)$.

For diffusive waves, $g > 1$, the coupling to the surroundings increases and modes overlap spectrally. The eigenstates are extended throughout the sample and the coupling to the modes from the left and the right sides are typically similar so that $\|W_{Ln}\|^2 \sim \|W_{Rn}\|^2$. Hence, $C_n \sim 1$ and the lower peak in $P(C_n)$ and $P(T_n)$ disappears. The probability distributions of $K_n$ and $T_n \sim K_n$ are then broad with peaks shifting towards values much greater than unity. Values of $T_n$ as large as 150 are found.

The variation of $\langle C_n \rangle$, $\langle K_n \rangle$ and $\langle T_n \rangle$ with $g$ are shown in Fig. 3(d). $\langle K_n \rangle$ and $\langle T_n \rangle$ first increase with $g$ as the correlation between eigenfunctions increases, but then decrease once the sample is translucent, $g \geq N/2$. The eigenfunctions in this regime are only slightly perturbed from the orthonormal eigenfunctions of the empty waveguide so the degree of nonorthogonality is small.

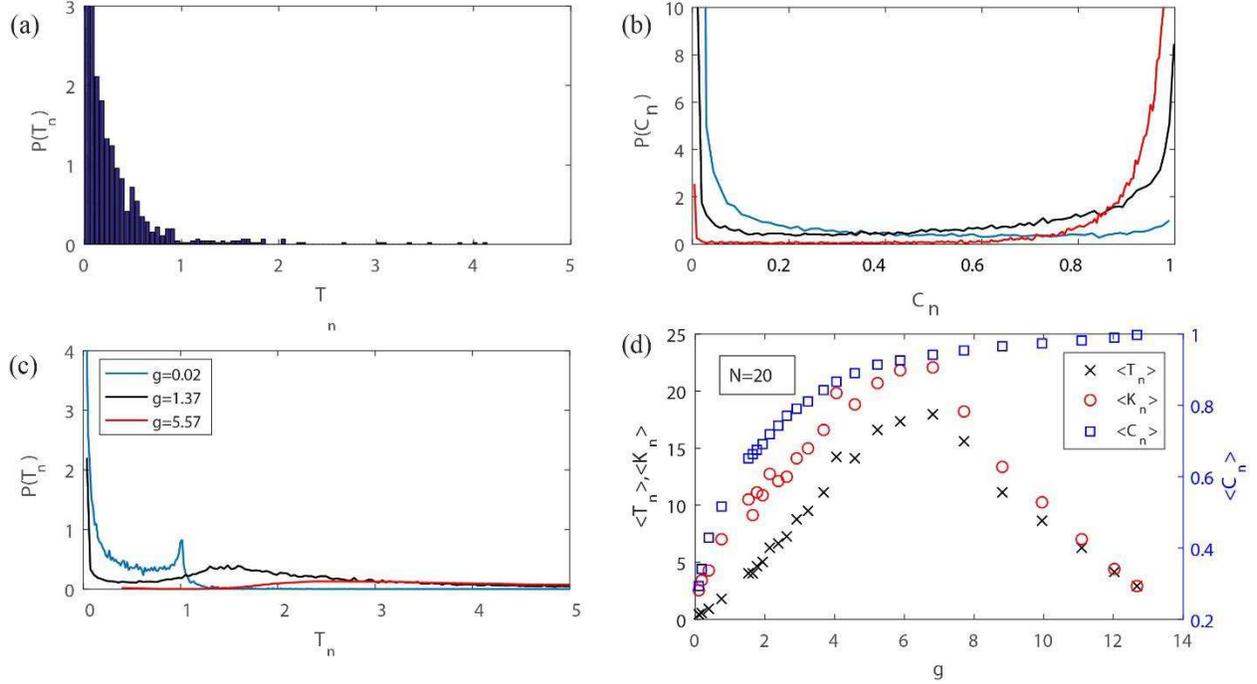

FIG. 3: (a) Distribution of experimental modal strengths $P(T_n)$ in the strong coupling regime with $\delta = 1.52$. (b,c) Simulations of $P(T_n)$ and $P(C_n)$ in the inset for samples with $g = 0.02$ (blue line), $g = 1.37$ (black line), $g = 5.57$ (red line). (d) Variation of $\langle K_n \rangle$, $\langle T_n \rangle$ and $\langle C_n \rangle$ with the conductance $g$.

The fluctuations and the scaling of these parameters in chaotic cavities as a function of modal overlap $\delta$ are discussed in Supplemental Material [46]. Statistics of modes in chaotic cavities are well described by random matrix theory (RMT) for which the internal Hamiltonian is modeled by a real symmetric matrix drawn from the Gaussian Orthogonal Ensemble. Strong enhancements of $T_n$ and $K_n$ are also observed as $\delta$ increases.

It is worthwhile to consider separately the contributions to the transmittance of the on- and off-diagonal terms. From the modal decomposition of the TM given in Eq. (1), the transmittance $T(\omega)$ can be expressed in terms of modal components as

$$T(\omega) = \Sigma_n T_n(\omega) + \Sigma_{n \neq m} \frac{\left[W_{Rm}^\dagger W_{Rn}\right]\left[W_{Lm}^\dagger W_{Ln}\right]}{(\omega - \widetilde{\omega}_n)(\omega - \widetilde{\omega}_m^*)} \quad (6)$$

A perturbative approach in the limit of small modal overlap [40] shows that $\langle K_n \rangle \equiv \langle O_{nn} \rangle$ increases as $\sim 1 + \delta^2$ [46]. $\delta$ modes contribute to transmission so that the incoherent sum of modal strengths $\langle T_{\text{inc}}(\omega) \rangle = \langle \Sigma_n T_n(\omega) \rangle$ is increased relative to $g = \langle T(\omega) \rangle$ by a term scaling as $\delta^3 \sim g^3$. Transmission is then reduced by destructive interference between correlated modal components with $n \neq m$ in Eq. (6) ensuring that transmission is bounded by unity. The contribution of off-diagonal terms to the average transmittance scales as $-\delta^3$, as expected. Numerical results from RMT simulations shown in Fig. 4 are in excellent agreement with the perturbative approach for $\delta < 1.4$.

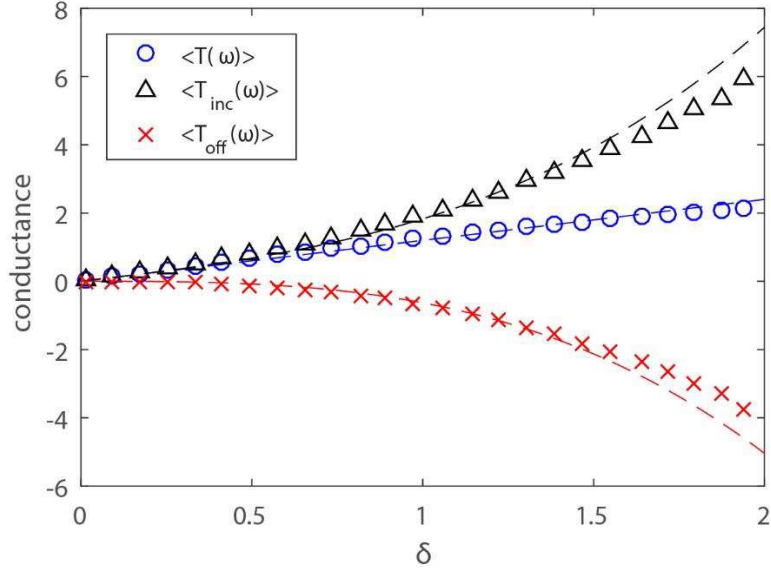

FIG. 4: Random matrix theory simulations of the scaling of the transmittance (blue circles), and the sum of the diagonal (black triangles) and off-diagonal (red crosses) modal contributions as a function of $\delta$. The dashed lines are the fits using analytical expression given in Supplementary Material [46].

The decomposition of the TM into its modal components provides a fresh vantage point from which to understand and control transport through and energy within disordered photonic and plasmonic media, chaotic cavities, and multimode fibers [32, 33]. The ability to approach perfect transmission in diffusive systems by exciting the first transmission eigenchannel and the maximal contrast in focusing through random systems via control of the TM increases with the number of resonant modes participating in transmission [34]. Thus the correlation in transmission eigenvalues is linked to the correlation in the modes of the medium. This is particularly important since modes are defined over the full spectrum, while eigenchannels are defined at a single frequency. The spectral correlation of modes may therefore be used to enhance control of transmission and delay times for broadband pulses [30, 52, 53].

**Acknowledgements**

This publication was supported by the European Union through the European Regional Development Fund (ERDF), by the French region of Brittany and Rennes Métropole through the CPER Project SOPHIE/STIC & Ondes, and by the National Science Foundation under grant DMR/-BSF: 1609218. We would like to acknowledge Jing Wang for discussions of early measurements of modal correlation and Zhou Shi, Ulrich Kühl, Olivier Legrand and Fabrice Mortessagne for stimulating discussions.

# Supplementary Material for 'Impact of non-Hemiticity on modal strength and correlation in transmission through random open cavities'

Matthieu Davy[1] and Azriel Z. Genack[2]

[1]Univ Rennes, CNRS, Institut d'Électronique et de Télécommunications de Rennes, UMR–6164, F-35000 Rennes, France
[2]Department of Physics, Queens College and Graduate Center of the City University of New York, Flushing, New York 11367, USA

## I. Demonstration of Eq. (2) of the main text

The statistical properties of eigenfunctions of open cavities can be found by considering the eigenvalues and eigenvectors of the effective non-Hermitian Hamiltonian, $H_{\text{eff}} = H_0 - \frac{i}{2} V V^T$. Here $H_0$ is the Hamiltonian of the closed system and $V$ is a real matrix which describes the coupling of external channels to the system. The matrix $\phi$ of eigenvectors of the effective Hamiltonian is defined by $H_{\text{eff}} \phi = \phi \widetilde{\Omega}$. Here $\widetilde{\Omega}$ is the diagonal matrix of eigenvalues of $H_{\text{eff}}$. We multiply this equation by $\phi^\dagger$ on the left so that

$$\phi^\dagger H_{\text{eff}} \phi = U \widetilde{\Omega}. \tag{S1}$$

Here $U = \phi^\dagger \phi$ is the Bell-Steinberger non-orthogonality matrix with elements $U_{mn} \equiv \phi_m^\dagger \phi_n$.[1-3] This gives the correlation between eigenfunctions and is related to the modal overlap matrix $O_{mn} \equiv U_{mn}^2$. Similarly

$$\phi^\dagger H_{\text{eff}}^\dagger \phi = \widetilde{\Omega}^\dagger U. \tag{S2}$$

Subtracting Eq. (S1) from (S2) and using $H_{\text{eff}} = H_0 - \frac{i}{2} V V^T$, we obtain $U \widetilde{\Omega} - \widetilde{\Omega}^\dagger U = \phi^\dagger V^T V \phi$.

To relate the modal overlap matrix to the decomposition of the scattering matrix (or the transmission matrix) into modal components, we use the vectors $W_n$ which give the projection of the eigenfunctions onto the channels, such as $W = V\phi$. The vectors $W_n$ concatenate the components associated to the modal speckle pattern on the left and right surfaces of the sample, respectively.

Equation (S2) leads to $U\widetilde{\Omega} - \widetilde{\Omega}^\dagger U = W^\dagger W$. Elements of the $U$ matrix can therefore be expressed in terms of the modal components of the scattering matrix as $U_{nm} = i \frac{W_n^\dagger W_m}{\widetilde{\omega}_m^* - \widetilde{\omega}_n}$. This finally leads to Eq. (2) of the main text using that $O_{mn} = U_{mn}^2$.

## II. Matrix of overlap

### A. Relation between transmission modal overlap matrix $\widetilde{O}(\widetilde{\omega}_1, \widetilde{\omega}_2)$ and the overlap matrix $O(\widetilde{\omega}_1, \widetilde{\omega}_2)$

The modal overlap matrix given in Eq. (2) of the main text is $O_{mn} = -\left(W_m^\dagger W_n\right)^2/(\widetilde{\omega}_n - \widetilde{\omega}_m^*)^2$.[4,5] The components of the $W_n$ vector associated with the left and right sides of the sample are $W_{Ln}$ and $W_{Rn}$, respectively, so that $W_n = [W_{Ln}\ W_{Rn}]$. This gives

$$O_{mn} = -\frac{\left(W_{Lm}^\dagger W_{Ln} + W_{Rm}^\dagger W_{Rn}\right)^2}{(\widetilde{\omega}_n - \widetilde{\omega}_m^*)^2} \tag{S3}$$

However, from measurements of the transmission matrix (TM), we can only extract the modal TMs (MTMs) $t_n = -i W_{Rn} W_{Ln}^T$. Hence, the vectors $W_{Ln}$ and $W_{Rn}$ cannot be found separately and the relative phase and amplitude between them cannot be determined from the TM. However, we can write

$$\left(W_{Lm}^\dagger W_{Ln} + W_{Rm}^\dagger W_{Rn}\right)^2 = \left(W_{Lm}^\dagger W_{Ln}\right)^2 + \left(W_{Rm}^\dagger W_{Rn}\right)^2 + 2\left(W_{Lm}^\dagger W_{Ln}\right)\left(W_{Rm}^\dagger W_{Rn}\right). \tag{S4}$$

When the disorder is uniform in space and when eigenfunctions extend throughout the sample, the average degree of correlation between modal components on the left is the same as for modal components on the right. The statistical properties of the vectors $W_{Ln}$ and $W_{Rn}$ are then equivalent so that $\|W_{Rn}\|^2 \sim \|W_{Ln}\|^2$. In the limit $N \gg 1$, we can approximate $W_{Lm}^\dagger W_{Ln} \sim W_{Rm}^\dagger W_{Rn}$ so that

$$\left(W_{Lm}^\dagger W_{Ln} + W_{Rm}^\dagger W_{Rn}\right)^2 \sim 4\left(W_{Lm}^\dagger W_{Ln}\right)\left(W_{Rm}^\dagger W_{Rn}\right). \tag{S5}$$

This can also be expressed in terms of the trace of the MTMs as, $\left(W_{Lm}^\dagger W_{Ln}\right)\left(W_{Rm}^\dagger W_{Rn}\right) = \text{Tr}\left(t_m^\dagger t_n\right)$. We therefore compute the transmission modal overlap matrix (see Eq. (3) of the main text)

$$\widetilde{O}(\widetilde{\omega}_m, \widetilde{\omega}_n) = -\frac{4\text{Tr}\left(t_m^\dagger t_n\right)}{(\widetilde{\omega}_n - \widetilde{\omega}_m^*)^2} \tag{S6}$$

The diagonal elements of $\widetilde{O}(\widetilde{\omega}_m, \widetilde{\omega}_n)$ give the modal strengths in transmission, $T_n = 4\|W_{Rn}\|^2 \|W_{Ln}\|^2 / |\Gamma_n|^2$. We show in the main text that $T_n$ can be written as the product $T_n = C_n K_n$. Here $C_n = 4\|W_{Rn}\|^2 \|W_{Ln}\|^2 / (\|W_{Rn}\|^2 + \|W_{Ln}\|^2)^2$ is the asymmetry in coupling of the nth eigenfunctions to the left and right boundaries. When $C_n \sim 1$, the degree of correlation between eigenfunctions is the same on the two sides so that the right hand sides of Eq. (S6) and Eq. (S3) are equal.

### B. Estimation of the homogenous losses

To estimate the linewidth $\Gamma_a$ associated with homogeneous absorption and losses through the antennas $\widetilde{\Gamma}_n$, related by $\Gamma_n = \widetilde{\Gamma}_n + \Gamma_a$, we observe that the average $\langle \widetilde{\Gamma}_n \rangle$ scales linearly with the number of coupled antennas, $\langle \widetilde{\Gamma}_n \rangle = 2N\langle \widetilde{\Gamma}_n^0 \rangle$, where $\langle \widetilde{\Gamma}_n^0 \rangle$ is the average linewidth associated to a single antenna coupled to the sample. By disconnecting the antennas from the switches, it is possible to decrease the number of coupled channels and thereby obtain an estimate for $\Gamma_a$. In the weak coupling regime, we find $\langle \widetilde{\Gamma}_n^0 \rangle \sim 0.15$ MHz and $\Gamma_a \sim 3.6$ MHz, so that the broadening of the resonances in the weak coupling regime is therefore mainly due to homogeneous absorption within the sample. In the strong coupling regime associated to a higher frequency range, $\langle \widetilde{\Gamma}_n^0 \rangle \sim 1$ MHz MHz and $\Gamma_a \sim 4$ MHz show that losses are dominated by the coupling through the antennas for $N = 8$.

## C. Effect of absorption on $T_n$

Uniform absorption broadens the linewidths (see Methods). The linewidths can be expressed as the sum $\Gamma_n = \Gamma_a + \tilde{\Gamma}_n$, where the $\Gamma_a$ is associated with the homogeneous absorption and $\tilde{\Gamma}_n$ results from the coupling of the system to the surroundings through the channels. $\tilde{\Gamma}_n$ is related to the coupling strength of the eigenfunction by $\sqrt{K_n}\tilde{\Gamma}_n = \|W_n\|^2$. The diagonal terms of the overlap matrix $\tilde{O}$ are therefore given by $T_n = \tilde{O}_{nn} = -\frac{4\|W_{Rn}\|^2 \|W_{Ln}\|^2}{(\tilde{\Gamma}_n+\Gamma_a)^2}$ and

$$T_n = T_n^0 \frac{\tilde{\Gamma}_n^2}{(\tilde{\Gamma}_n+\Gamma_a)^2} \tag{S7}$$

Here $T_n^0 = -\frac{4\|W_{Rn}\|^2 \|W_{Ln}\|^2}{\tilde{\Gamma}_n^2}$ is the modal transmission in the absence of absorption. Equation (S7) demonstrates that $T_n$ is lowered by absorption.

## III. Simulation results on the negative correlations between eigenfunctions

In order to look at the average modal selectivity in a large number of samples and confirm the negative correlation between eigenfunctions of overlapping resonances found in microwave measurements, we carry out simulations utilizing the recursive Green's function method[6] in random quasi-1D samples. A waveguide is connected to leads supporting $N$ channels at the left and right interfaces and the disorder is statistically uniform within the system. Details of the simulation method can be found in Ref. [7].

For three ensemble with different conductance $g$, we find spectra of the TM $t(\omega)$ and perform the modal analysis The number of channels and conductance are equal to $N = 16$ for $g = 0.106$ and $g = 0.64$, and $N = 33$ for $g = 1.03$.

The modal overlap matrix given by Eq. (3) of the main text is then computed and the average of $\langle \tilde{O}(\tilde{\omega}_1, \tilde{\omega}_2) \rangle$ are shown in Supplementary Figure S1 for the three ensembles. As for experimental values, $\langle \tilde{O}(\tilde{\omega}_1, \tilde{\omega}_2) \rangle$ are presented as a function of the complex shift between two resonances $|\tilde{\omega}_1 - \tilde{\omega}_2|$ normalized by the average level spacing $\Delta\omega$. We first observe that the magnitude of negative correlation increase with the conductance $g$, as expected. As the modal overlap $\delta \sim g$ increases, the degree of non-Hermiticity of the system also increases so that the degree of negative correlation between eigenfunctions with the same complex spacing is enhanced.

The simulation results presented in Supplementary Fig. S1 are in good agreement with Eq. (4) of the main text when the complex spacing between two resonances is normalized by the level spacing and a scale factor $a$ of order of unity, $\delta\tilde{\omega} = (\tilde{\omega}_1 - \tilde{\omega}_2)/(a\Delta\omega)$, such as, $O(\tilde{\omega}_1, \tilde{\omega}_2) \sim -\frac{1}{|\delta\tilde{\omega}|^4}[1 - (1 + |\delta\tilde{\omega}|^2)\exp(-|\delta\tilde{\omega}|^2)]$. This confirms experimental observations.

We find that the scale factor is the same for the two configurations with $N = 16$ and the same length. We may therefore hypothesize that $a$ only depend on the dimension of the medium. This is consistent with results of random matrix theory which show that the statistics of eigenvectors are universal when the complex spacing between eigenvalues $\widetilde{\omega}_m$ and $\widetilde{\omega}_n$ of $M$x$M$ non-Hermitian random matrices is normalized such as $\delta\widetilde{\omega} = \sqrt{M}(\widetilde{\omega}_m - \widetilde{\omega}_n)$. [8,9]

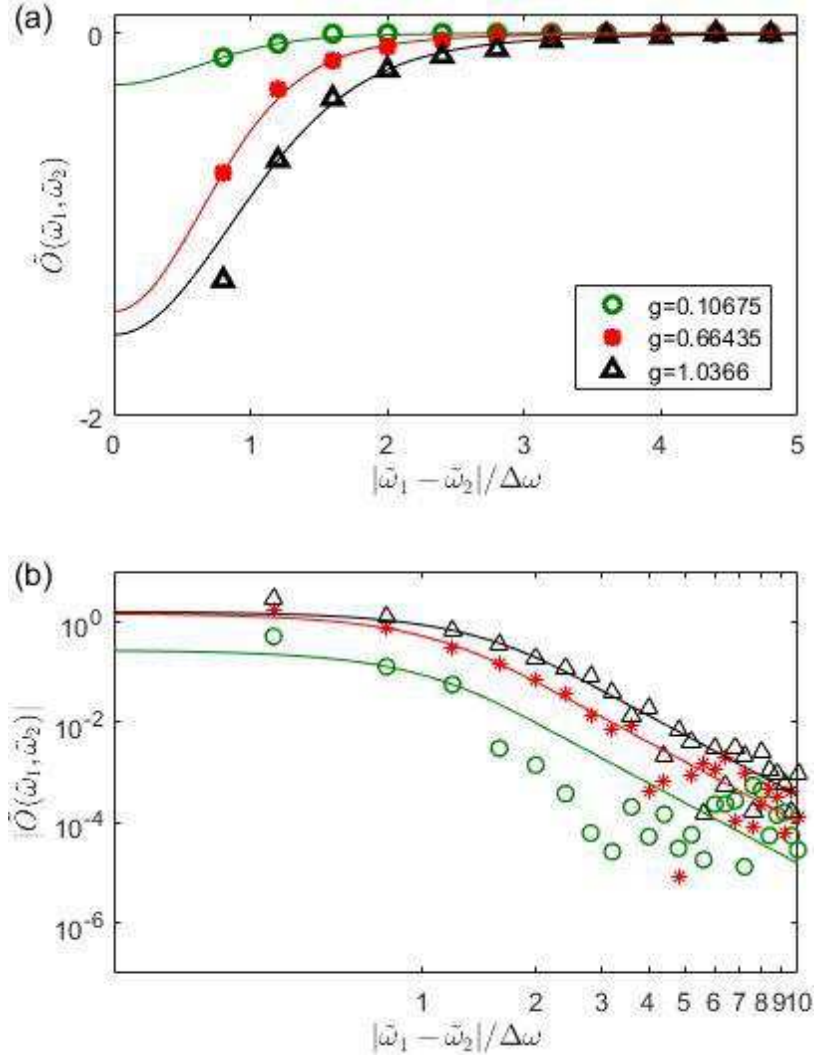

**Supplementary Figure S1: Distribution of the correlator $\widetilde{O}(\widetilde{\omega}_1, \widetilde{\omega}_2)$ in simulations.** $\widetilde{O}(\widetilde{\omega}_1, \widetilde{\omega}_2)$ is plotted as a function of $|\delta\widetilde{\omega}| = |\widetilde{\omega}_1 - \widetilde{\omega}_2|$ normalized by the average spacing between modes $\Delta\omega$ and a scale factor $a = 0.7$, $a = 0.73$ and $a = 0.9$ for the three ensembles $g = 0.106$ and $g = 0.64$, and $N = 33$ for $g = 1.03$. The results are presented in linear (a) and logarithmic (b) scales. The lines are fits to theoretical prediction given in Eq. (4) of the main text.

### IV. Distribution of modal strengths in transmission for chaotic cavities

#### A. Random Matrix Theory (RMT) simulations

In chaotic cavities, the eigenfunctions extend throughout the sample even though the coupling to channels may be weak. Statistics of modes in chaotic cavities are well described by random matrix theory (RMT).

The $M \times M$ internal Hamiltonian $H_0$ of the effective Hamiltonian $H_{\text{eff}} = H_0 - \frac{i}{2}\gamma V V^T$ is modeled by a real symmetric matrix drawn from the Gaussian Orthogonal Ensemble with $\langle (H_0)_{ij}^2 \rangle = 1/M$. The coupling matrix $V$ is a real random matrix with Gaussian distribution and $\langle V_{ij}V_{kl}\rangle = \delta_{ij}\delta_{kl}/M$. The modal overlap increases with increasing coupling strength of the channels to the system, $\kappa = \gamma/2$. The coupling strength to the continuum is $\kappa_c = \pi\gamma/(2MD)$, where $D = \pi/M$ is the mean level spacing at the center of the band, $E = 0$. The coupling strength is related to the transmission coefficient through the channels, $T_c$, with $T_c = 4\kappa/(1+\kappa)^2$.

### B. Distribution of asymmetry factor $C_n$

We first express the asymmetry between projections of the eigenfunctions on the incoming and outgoing channel, $C_n$, as

$$C_n = 1 - \left(\frac{\|W_{Ln}\|^2 - \|W_{Rn}\|^2}{\|W_{Ln}\|^2 + \|W_{Rn}\|^2}\right)^2 \tag{S8}$$

Since the eigenfunctions are uniformly distributed over the volume for chaotic systems, the terms $\|W_{Ln}\|^2$ and $\|W_{Rn}\|^2$ are the sum of $N$ independent random variables. The average of $C_n$ depends on $N$ and the degree of complexness of the eigenfunctions defined as $q_n^2 = \langle \text{Im}(\phi_n)^2 \rangle / \langle \text{Re}(\phi_n)^2 \rangle$. $q_n^2$ is related to the phase rigidity $\rho_n$ of the eigenfunctions and the Petermann factor, $K_n = \rho_n^{-2}$, with $q_n^2 = (1-\rho_n)/(1+\rho_n)$.

For a given $q_n$ in the limit $N \gg 1$, Eq. (S8) yields $C_n \sim (1+q_n)N/(1+(1+q_n)N)$. When the modal coupling vectors $W_{Ln}$ and $W_{Rn}$ are real, which is the case of the weak coupling regime, $q_n \sim 0$ gives $C_n = N/(N+1)$. However, for complex coupling vectors with statistically equivalent real and imaginary parts, $q_n \sim 1$, and $C_n = 2N/(2N+1)$.

The term $(\|W_{Ln}\|^2 - \|W_{Rn}\|^2)/(\|W_{Ln}\|^2 + \|W_{Rn}\|^2)$ is mainly the difference of two independent random variables $\|W_{Ln}\|^2$ and $\|W_{Rn}\|^2$. In the limit $N \gg 1$, this is a Gaussian variable. $1 - C_n$ is the square of a Gaussian variable and therefore has the Porter-Thomas distribution[14]

$$P(C_n; q_n) = \sqrt{\frac{1}{2\pi(1-C_n)[1+(1+q_n)N]}} \exp\left(-\frac{[1+(1+q_n)N]}{2}\right) \exp\left(-\frac{(1-C_n)}{2}\right) \tag{S9}$$

The distribution of $C_n$ is finally given by $P(C_n) = \int P(C_n, q_n) P(q_n) dq_n$. This however requires the distribution of $q_n$, which is known analytically only in the weak coupling regime defined by $\frac{\sqrt{\text{Var}(\Gamma_n)}}{\Delta} \ll 1$. Nevertheless, in the limit $N \gg 1$, fluctuations in $q_n$ are small so that we can make the approximation $q_n \sim \langle q_n \rangle$. Numerical results shown in Supplementary Fig. S2a are in good agreement with the analytical result with $q_n = \langle q_n \rangle$ for $\kappa = 1$, but deviations are observed for small values of $N$ due to fluctuations in $q_n$.

### C. Distribution of $K_n$ and $T_n$

The distribution of the Petermann factor $P(K_n)$ and modal strength $P(T_n)$ are shown in Supplementary Figs. S2b,c. In the weak coupling regime ($\kappa \ll 1$), most resonances are isolated so that $P(K_n)$ and $P(T_n)$ are both peaked near $T_n = 1$. Values of $T_n$ smaller than unity are a consequence of the asymmetry of modes for which $C_n < 1$. As $\kappa$ increases, the two distributions broaden. Long tails are observed for $P(K_n)$ and $P(T_n)$ for $\kappa = 0.4$ and $\kappa = 1$. Values of the Petermann factor and modal strengths as high as 300 are found.

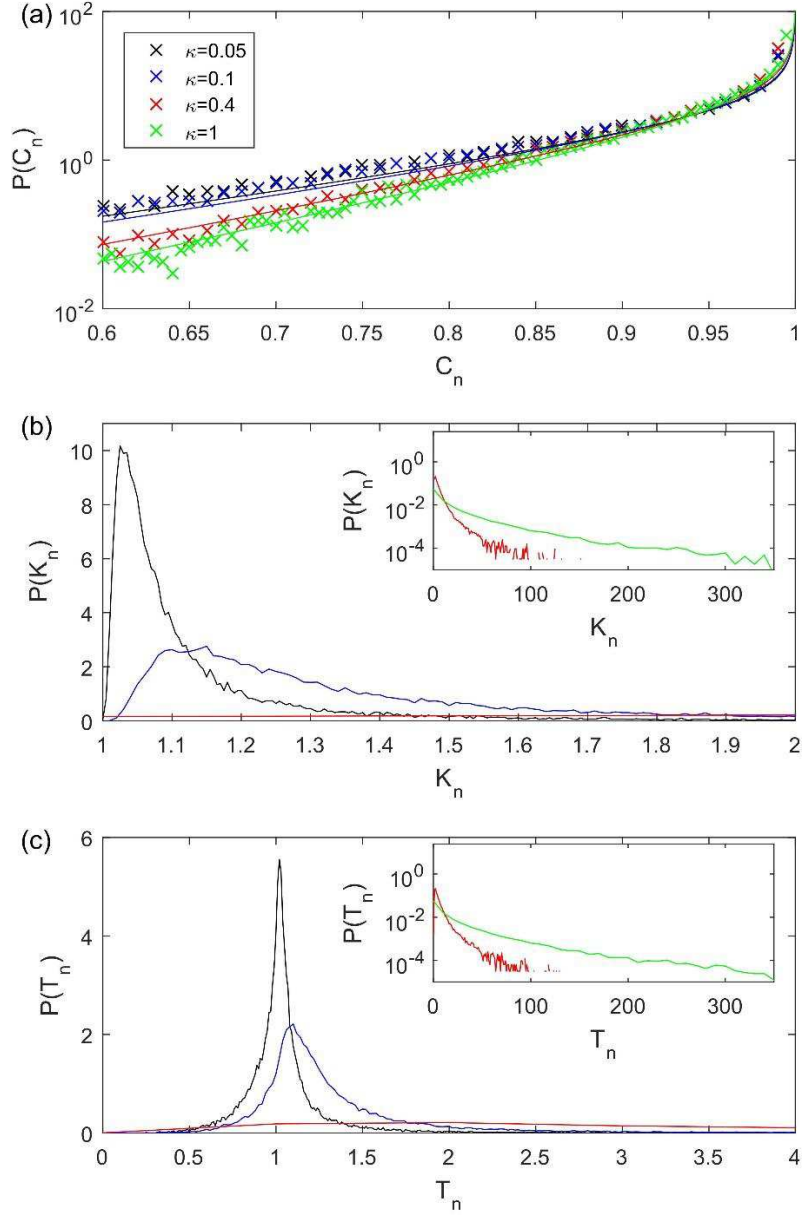

**Supplementary Figure S2: Probability distributions modal characteristics.** The probability distributions of the asymmetry factor, $C_n$ (a), the Petermann factor, $K_n$ (b) and the modal strength in transmission for chaotic cavities $T_n$ (c) are shown in random matrix simulations with $2N = 20$ coupled channels, for different values of the coupling parameter $\kappa$, $\kappa = 1$ (in green) $\kappa = 0.4$ (in red), $\kappa = 0.1$ (in blue) and $\kappa = 0.05$ (in black). $P(K_n)$ and $P(T_n)$ for $\kappa = 0.4$ and $\kappa = 1$ are presented in a semilog scale in the inset of (b) and (c) to observe that the distributions reach the large values in the tail of the distribution.

### V. Scaling of diagonal and off-diagonal terms sums modal contributions to transmission

Using the expression for the transmittance $T(\omega)$ in terms of modal components

$$T(\omega) = \Sigma_n T_n(\omega) + \Sigma_{\substack{n,m \\ n \neq m}} \frac{[W_{Rm}^\dagger W_{Rn}][W_{Lm}^\dagger W_{Ln}]}{(\omega-\widetilde{\omega}_n)(\omega-\widetilde{\omega}_m^*)}, \tag{S10}$$

we evaluate the scaling of the diagonal (incoherent) modal sum, $\langle T_{\text{inc}}(\omega)\rangle = \langle \Sigma_n T_n(\omega)\rangle$ and the off-diagonal sum $\langle T_{\text{off}}(\omega)\rangle = \langle T(\omega) - \langle T_{\text{inc}}(\omega)\rangle\rangle$, as a function of modal overlap $\delta = \langle\Gamma\rangle/\Delta\omega$ in the weak coupling regime. $\delta$ gives the number of modes contributing to transmission so that the incoherent sum is $\langle T_{\text{inc}}(\omega)\rangle \sim a_1 \delta \langle K_n \rangle$, where $a_1 \sim 1$ is near unity. To obtain the scaling of the average of the Petermann factor $\langle K_n \rangle$, we express the eigenvectors $|\phi_n\rangle$ of the effective Hamiltonian for a cavity, $H_{\text{eff}} = H_0 - \frac{i}{2}VV^T$, in the basis $\{|\psi_n\rangle\}$ of the unperturbed eigenvectors of the Hamiltonian of the closed system $H_0$ [15]

$$\phi_n = \frac{1}{\sqrt{1-q_n^2}}\left(\psi_n - i\Sigma_{p\neq n}\frac{\Gamma_{np}}{2(\omega_n-\omega_m)}\psi_p\right) \tag{S11}$$

The degree of complexness of an eigenfunction is $q_n^2 = \Sigma_{p\neq n}\frac{\Gamma_{np}^2}{4(\omega_n-\omega_m)^2}$ with $\Gamma_{np} = \Sigma_c V_n^c V_p^c$. In the weak coupling regime, $q_n^2 \ll 1$ and $K_n \sim 1 + 4q_n^2$. Using the expression for $\langle q_n^2\rangle$ given in Ref. [15] in the limit $N \gg 1$, $\langle q_n^2\rangle = f\delta^2/M$, we obtain $\langle K_n\rangle = 1 + \frac{4f\delta^2}{M}$. Here, $f = \langle \Sigma_{p\neq n}\Delta^2/[4(\omega_n-\omega_p)^2]\rangle$ is a factor that depends solely on the statistics of the central frequencies of resonances of the closed system. For instance, equally spaced central resonances, which occur in the Picket-fence model, gives $f \sim \pi^2/12$ [15]. An expression for $\langle T_{\text{inc}}(\omega)\rangle$ is then

$$\langle T_{\text{inc}}(\omega)\rangle = a_1\delta(1 + \frac{4f\delta^2}{M}). \tag{S12}$$

The enhancement of the incoherent sum of modes implies that transmission is reduced by destructive interference between the modes since transmission is bounded by unity. Moreover the average transmission involving the excitation of all modes in a random medium is proportional to the ratio of the mean free path $\ell$ and the sample length $L$, $\langle T_a\rangle \sim \ell/L$. This is much smaller than unity for strongly multiply scattering media in which $\ell \ll L$. Destructive interference is a result of the correlation between modal field speckle patterns, $\left(W_{Rm}^\dagger W_{Rn}\right)\left(W_{Lm}^\dagger W_{Ln}\right)$ in the sum over all off-diagonal elements $T_{\text{off}}(\omega) = \Sigma_{n\neq m}\frac{[W_{Rm}^\dagger W_{Rn}][W_{Lm}^\dagger W_{Ln}]}{(\omega-\widetilde{\omega}_n)(\omega-\widetilde{\omega}_m^*)}$.

Since the vectors $W_{Ln}$ and $W_{Rn}$ are independent, in the limit $N \gg 1$, we can make the approximation $\left(W_{Rm}^\dagger W_{Rn}\right)\left(W_{Lm}^\dagger W_{Ln}\right) \sim \left(W_m^\dagger W_n\right)^2/4$ as shown in Section IIA. Using Eq. (S9), we obtain

$$T_{\text{off}}(\omega) = -\Sigma_{\substack{n,m \\ n\neq m}}\frac{O_{mn}(\widetilde{\omega}_m^* - \widetilde{\omega}_n)^2}{4(\omega-\widetilde{\omega}_n)(\omega-\widetilde{\omega}_m^*)} \tag{S13}$$

To approximate $\langle T_{\text{off}}(\omega)\rangle$, we first use the completeness relation of eigenfunctions, which yields $\Sigma_{m\neq n}O_{nm} = 1 - K_n$. The sum can be restricted to $\delta$ strongly overlapping modes since the degree of correlation of the eigenfunctions $O_{nm}$ falls rapidly with frequency shift for modes separated by more than a linewidth. Using the same idea, we can make the approximation, $\frac{(\widetilde{\omega}_m^* - \widetilde{\omega}_n)^2}{4(\omega-\widetilde{\omega}_n)(\omega-\widetilde{\omega}_m^*)} \sim -\frac{(\Gamma_m+\Gamma_n)^2}{4\Gamma_n\Gamma_m}$. In the limit $N \gg 1$ in diffusive systems, fluctuations in the linewidths are small relative to the average linewidth [16], so that $\Gamma_n \sim \langle\Gamma_n\rangle$ for $N \gg 1$. This gives

$$\langle T_{\text{off}}(\omega)\rangle \sim a_1\delta(1-\langle K_n\rangle) = -\frac{4af\delta^3}{M}. \quad (S14)$$

The sum of the diagonal and off-diagonal terms satisfies the relation $\langle T\rangle = \langle T_{\text{inc}}(\omega)\rangle + \langle T_{\text{off}}(\omega)\rangle = a_1\delta$, with $a_1$ of order 1, in agreement with the Thouless relation in random media $g = \delta$.